\documentclass[12pt]{article}
\usepackage{amsmath, amssymb, amscd}
\usepackage{graphicx}

\begin{document}
\date{}

\title{Models of Opinion Formation: Influence of Opinion Leaders}

\author{\textsc{Nino Boccara}\\
Department of Physics, University of Illinois, Chicago, USA\\
and\\
DRECAM/SPEC, CE Saclay, 91191 Gif-sur-Yvette Cedex, France\\
\texttt{boccara@uic.edu}}

\maketitle

\section*{Abstract}
This paper studies the evolution of the distribution of opinions in a population of individuals in which there exist two distinct subgroups of highly-committed, well-connected opinion leaders endowed with a strong convincing power. Each individual, located at a vertex of a directed graph, is characterized by her name, the list of people she is interacting with, her level of awareness, and her opinion. Various temporal evolutions according to different local rules are compared in order to find under which conditions the formation of strongly polarized subgroups, each adopting the opinion of one of the two groups of opinion leaders, is favored.

\section{Introduction}
Social choice theory, as a scientific discipline, started with the works of the eigthteenth-century french mathematician Jean-Charles de Borda (1733-1799) who showed that majority rules could lead to inconsistent results when voters have to choose between three candidates. But it is only in 1951 when Kenneth Arrow published his celebrated impossibility theorem~\cite{ka1951}, showing that there exists no social choice rule satisfying a set of reasonable requirements, that social choice became a scientific discipline.

The purpose of this paper is to describe and study models of opinion formation under the influence of opinion leaders favoring the existence of polarized subgroups of individuals.  Although in a society each individual has personal opinions,  these opinions could, to a certain extent,  change under the influence of  the opinions of the group of individuals this person is connected to, such as family members, friends, coworkers, and other persons who by virtue of position may exercise influence like politicians or journalists.  Since these individuals, influenced by other individuals, may also revise their own opinions, we observe successive modifications of each individual's opinions; and one may reasonably ask if this iterative process leads to the formation of consensual subgroups of people. 

In democratic societies, the outcome of political elections plays an essential role and it is not surprising to discover that the problem of voter decision-making as attracted many political scientists. In 1992, John Zaller, in a book~\cite{jz1992}, considered as a most important contribution to political science~\cite{ gm1992, hk1993}, developed a theory to explain how people receive political information and determine their political preferences. Following Converse~\cite{pc1962} Zaller argues that most people do not have fixed positions on issues, only the most aware individuals, who are well informed, have a consistent ideology. In his own words ``there is high variance in political awareness around a generally low mean.'' Public opinion is shaped by exposure to elite discourse, via the media, on issues with, however, significant differences in attention to this  discourse. ``Political awareness denotes intellectual or cognitive engagement with public affairs as against emotional or affective engagement or no engagement at all.'' The most aware individuals are more able to receive political information but, due to their exposure to multiple and often conflicting messages, are more selective in accepting ideas contradicting their basic values. The least aware individuals receive less information and are usually more likely to be influenced. Thus, variations of political opinions and political awareness are strongly correlated. In our model, described below, the dependence of an individual's awareness on her opinion plays an important role.

The construction of a mathematical model of opinion dynamics implicitly assumes that opinions can be measured. In the 1920s it was controversial among psychologists whether attitudes could be measured. In 1928, Louis Leon Thurstone showed that it could be done~\cite{t1928} and even led to important results~\cite{pt1933}.

Recently quite a few number of agent-based models of opinion formation have been studied~\cite{s-ws2000, s2001, bcas2001, wdan2002, hk2002}. In all these models as in ours, each individual is represented by an agent whose state is a simplified description of the individual's characteristics.

In this paper we define and study different versions of a model of opinion formation. We represent a society of individuals by a directed graph, that is an ordered pair of disjoint sets $(V,E)$, where $V$ is a nonempty set of elements called vertices, nodes, or points, and $E$ a set of ordered pairs of distinct elements of $V$, called  directed edges, arcs, or links.  Each vertex is occupied by an individual whose state describes her name, the list of individuals she interacts with, her level of awareness, and her opinion. The opinion of an individual located at vertex $i$ evolves as a result of her interactions with the individuals located at all the vertices $j\in V$ such that  $(i,j)\in E$, according to an evolution rule that takes into account opinions and levels of awareness of the interacting individuals. All the individuals directly connected to a given individual $i$ form her \emph{social environment} or \emph{social neighborhood}. In what follows, the words \emph{neighbor} and \emph{neighborhood} will not imply any spatial proximity; a neighbor of $i$ is just an element of her social environment who may influence her opinion.

\section{The Model}
Our model of opinion formation in a population of individuals is a scale-free social network in which each vertex is occupied by an individual. Social networks, such as scientific collaboration networks~\cite{bjnrsv2002},  coauthorship networks of scientific papers~\cite{r1998},  human sexual contacts networks~\cite{leasa2001}, e-mail addresses networks~\cite{emb2002}, are characterized by a small average shortest path length between two randomly selected vertices, a high clustering coefficient and, in most cases, a power-law probability distribution for the vertices' degrees.~\cite{ab2002, nb2004}.

Each individual's level of awareness is represented by a real number $s$ between 0 and 1.  In our model, the opinion of an individual with a high $s$-value are thought to have more value;  she has a strong convincing power when interacting with other individuals and a high degree of ``wise'' skepticism when influenced by other individuals.  Concepts similar to our level of awareness have been introduced by various authors. French~\cite{f1956}  defines the \emph{power} of A over B (with respect to a given opinion) as the maximum force which A can induce on B minus the maximum resisting force which B can mobilize in the opposite directions; De Groot~\cite{dg1974}, Krause~\cite{k2000}, and Hegselmann and Krause~\cite{hk2002} define a stochastic matrix of positive coefficients  $a_{ij}$ representing the \emph{weight given by $i$ to $j$} that may change with time and opinions; Nowak {\it et al.}~\cite{nsl1990} in the numerical study of a model inspired by Latan\'e's theory of social impact~\cite{l1981}, among other attributes, characterize each individual by a strength variable called  \emph{persuasiveness}. In our model, the level of awareness of an individual is a time-independent characteristic of the individual. 

\begin{figure}[h]
\centering\includegraphics[scale=0.9] {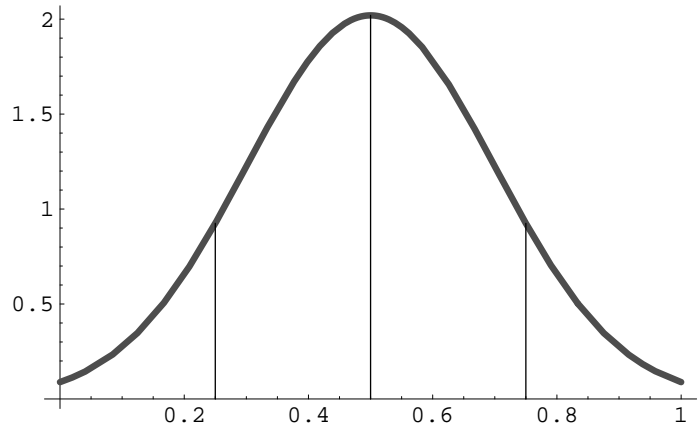}
\caption{\label{fig:awarenessPDF}\textit{Probability density function of the distribution of levels of awareness in the population. The vertical lines delimit four different groups of individuals (see text). }}
\end{figure}

For the probability distribution of levels of awareness in the society we have chosen a bell-shaped probability density function $p$ symmetric about $0.5$ given by 
\begin{equation}
p: s\mapsto p(s) = \frac{1}{2\pi \sigma n} \exp\frac{(s-0.5)^2}{2\sigma^2}
\label{eqn:pdf}
\end{equation}
where
$$
 \sigma = 0.2\  \text{and}\  n = \int_0^1\exp\frac{(x-0.5)^2}{2\sigma^2}\, dx = 0.9876.
$$
Choosing  $\sigma=0.2$ makes the percentages of the population belonging to the four groups  of individuals having, respectively, a level of awareness in the semi-open intervals $[0, 0.25[$, $[0.25, 0.5[$, $[0.5, 0.75[$, and $[0.75, 1[$ approximately equal to 10 \%, 40 \%, 40 \%, and 10 \% (exactly 10.07, 39.93, 39.93, and 10.07). The graph of $p$ is represented in Figure~\ref{fig:awarenessPDF}.

Each individual has an opinion represented by a real number in the interval $]0,1[$ uniformly distributed among all individuals. An individual's opinion may change at each time step according to three different probabilistic evolution rules: 
\begin{itemize}
\item \emph{1- Adopt leader's opinion.} The individual $i$ with a level of awareness equal to $s_i$ changes her opinion $\omega_i$ with the probability $(1-s_i)^\alpha$, where $\alpha$ is a positive real which is the same for all individuals, to adopt the opinion of the individual in her neighborhood who has the highest level of awareness if this level of awareness is higher than her own.
\item   \emph{2- Adopt weighted average opinion.} The individual $i$ with a level of awareness equal to $s_i$ changes her opinion $\omega_i$ with the probability $(1-s_i)^\alpha$, where $\alpha$ is a positive real which is the same for all individuals, to adopt the weighted average opinion of her neighbors $\overline{\omega_i}$ defined by
$$
\overline{\omega_i} = \frac{\displaystyle\sum_{k\in\textrm{neighborhood of $i$}} s_k\omega_k}  
 {\displaystyle\sum_{k\in\textrm{neighborhood of $i$}} s_k},
$$
where $s_k$ is the level of awareness of neighbor $k$ and $\omega_k$ her opinion, the sum being extended to all vertices $k$ such that $(i,k)$ is a directed link. 
\item   \emph{3- Adopt weighted average of neighbors' opinions having a higher level of awareness} Same as above except that an individual $i$ with a level of awareness equal to $s_i$ changes her opinion $\omega_i$ with the probability $(1-s_i)^\alpha$ to adopt the weighted average opinion of the subgroup of neighbors having a higher level of awareness than her own. This last rule is a sort of compromise between the two previous ones.
\end{itemize}

We also considered slightly different versions of the evolution rules just described in which the probabilistic condition is replaced by a threshold condition. That is, an individual $i$, instead of adopting a new opinion with a probability equal to $(1-s_i)^4$, adopts a new opinion if, and only if, the new opinion does not differ from her own by more that  $(1-s_i)^\alpha$, where  $\alpha$, as mentioned above, is the same positive real for all individuals. 

Note that for both probabilistic and threshold conditions, the exponent $\alpha$ characterizes the level of skepticism of the society as a whole. Increasing its value increases the level of skepticism of the whole population and makes it more difficult to convince. The exponent $\alpha$ can therefore be viewed as a cultural trait of the society. In our numerical simulations we took $\alpha = 4$.

Note also that the expressions of both versions of the evolution rules show that individuals with a low level of awareness are much easier to convince than individuals with a high level of awareness. The least aware individuals, that is, in our model, the individuals with a small level of awareness are responsible of the observed fickleness of voters in polls during election campaigns. 

To sum up, in our models, a social network model is represented by a directed graph with $N$ nodes. Each node is occupied by an individual characterized by
\begin{itemize}
\item[1-] her \emph{name}: an integer between $1$ and $N$;
\item[2-] her \emph{social environment}: a list of the individuals (called neighbors) she interacts with;
\item[3-] her \emph{level of awareness $s$}: a real in the semi-open interval $[0,1[$;
\item[4-] her \emph{opinion $\omega$}: a real in the open interval $]0,1[$.
\end{itemize} 

The size of each neighborhood is a random integer uniformly distributed between $1$ and a maximum value $n_{\max} = 7$ (implying an average number of neighbors equal to 4). An individual $k$ belongs to the neighborhood of individual $i\neq k$ if the directed link $(i,k)$ belongs to the set $E$ of graph's edges.  While vertices out-degrees are uniformly distributed between $1$ and $n_{\max}=7$, the random selection of neighbors is such that vertices in-degrees have a Pareto probability distribution with a minimum value parameter $x_0=1$ and a shape parameter $\sigma=2$. The probability distribution of in-degrees has, therefore, a cumulative distribution function given by $x\mapsto 1-x^{-2}$, and a density function equal to $x\mapsto 2x^{-3}$.\footnote{The cumulative distribution function of the Pareto distribution with a minimum value parameter equal to $x_0$ and a shape parameter equal to $\sigma$ is
$$
x\mapsto 1-\left(\frac{x_0}{x}\right)^\sigma.
$$}
 
To complete the definition of the model, we define a subgroup of individuals called \emph{opinion leaders}. This subgroup consists of the  $r N$ individuals located at the nodes having the highest in-degrees, that is, the individuals who belong to more neighborhoods. More precisely, each individual of the fraction $r$ of the total number $N$ of individuals, selected among those having the highest in-degrees, is assigned a level of awareness $s = 1$ and, with a probability $\frac{1}{2}$, an opinion either equal to $\Omega$ or $1-\Omega$.  This subnetwork of opinion leaders is, therefore, the group of the most influential individuals. Since their level of awareness is equal to 1, they never change opinion during the temporal evolution of the system. They represent a group of commmitted individuals  firmly attached to their opinion.
 
For example, in a population of 1000 individuals, with $\Omega=0.4$, the states of individuals 587 and  723 could be given by
\begin{eqnarray*}
&\{587,\  \{426, 802, 977, 679\},\  0.604395,\  0.543928\}\\
&\{723,\  \{770, 91, 367, 313, 967, 349, 68\},\  1.0,\  0.6\}.
\end{eqnarray*}
Individual 587 has 4 neighbors---individuals 426, 802, 977, and 679---, a level of awareness $s_{587}=0.604395$, and an opinion $\omega_{587} = 0.543928$, whereas individual 723 is an opinion leader who has 7 neighbors---individuals 770, 91, 367, 313, 967, 349, and 68---,  a level of awareness $s_{723}=1.0$, and an opinion $\omega_{587} = 1-\Omega = 0.6$.

At each time step, for all integers $i\in\{1, 2, \ldots, N\}$, the opinion $\omega_i$ of individual $i$ evolves according to one of the six evolution rules mentioned above. In a more general model we could consider that each individual has various opinions on different issues and apply similar evolution rules to each opinion.

\section{Numerical results}

\begin{figure}[h]
\centering\includegraphics[scale=0.9] {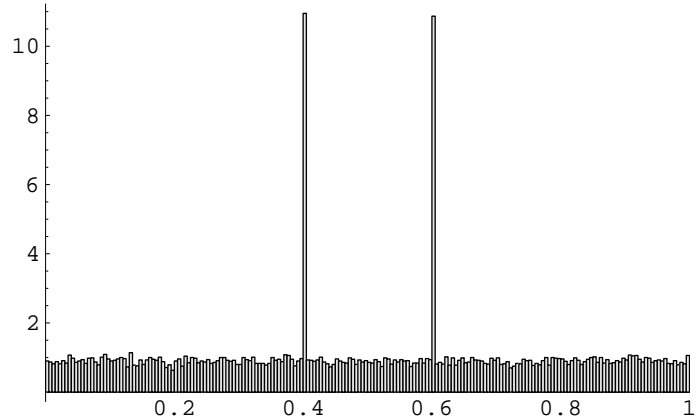}
\caption{\label{fig:initialhistogram}\textit{Histogram of the distribution of opinions averaged over the 20 initial societies.}}
\end{figure}

We performed several runs of the six different evolution rules applied to 20 different initial societies of 1000 individuals having a fraction $r=0.1$ of committed individuals with a level of awareness equal to 1.\footnote{In order to check our results we also made a few numerical simulations on larger societies of 10,000 individuals.}  These individuals have, with a probability $\frac{1}{2}$, an opinion either equal to $\Omega = 0.4$ or to $1-\Omega=0.6$. The levels of awareness of the remaining part of the population are randomly distributed following the probability density function $p$ given by~(\ref{eqn:pdf}). According to whether her level of awareness belongs to one of the following semi-open intervals: $[0,0.25[$, $[0.25,0.5[$, $[0.5, 0.75[$, or $[0.75, 1[$, we say that the individual belongs to group 1, 2, 3 or 4. 
Following Zaller, chosing $\alpha = 4$ implies that individuals in group 1 have a particularly low level of awareness and are, therefore, very easy  to convince, whereas individuals in group 4 have a high level of awareness and are especially difficult to convince. Groups 2 and 3 represent the majority; they have a moderate level of awareness and are more or less easy to convince according to whether they belong to group 2 or 3. The  average distribution of opinions in the 20 initial societies is represented in 
Figure~\ref{fig:initialhistogram} and the percentages of individuals belonging to the different groups are given in Table~\ref{initialNumbers}. In order to compare the different evolution rules, all the reported numerical results were obtained starting from the same 20 initial populations of 1000 individuals. 

\begin{table}[h]
 \begin{center}
 \begin{tabular}{|c|c|c|c|c|c|}
\hline
 group 1 & group  2 & group  3 &  group 4 & group 0.4 &  group 0.6\\ 
 \hline
 8.87        & 36.27        & 35.63       & 9.23  & 5.02  &  4.98\\ 
 \hline
\end{tabular}
\end{center}
\caption{\textit{Percentages of the total population belonging to each group; groups 0.4 and 0.6 denote the groups of strongly committed individuals having a level of awareness equal to 1 and an opinion either equal to 0.4 or 0.6.}}
\label{initialNumbers}
\end{table}

\subsection{Rule 1: Adopt leader's opinion with a probabilistic condition}

Starting from the 20 different initial populations of 1000 individuals described above, we studied their temporal evolutions assuming that each individual $i$ with level of awareness $s_i$ adopts the opinion of her neighbor having the highest level of awareness with a probability $(1-s_i)^4$ if, and only if, this level of awareness is higher than her own. 

\begin{figure}[h]
\centering\includegraphics[scale=0.9] {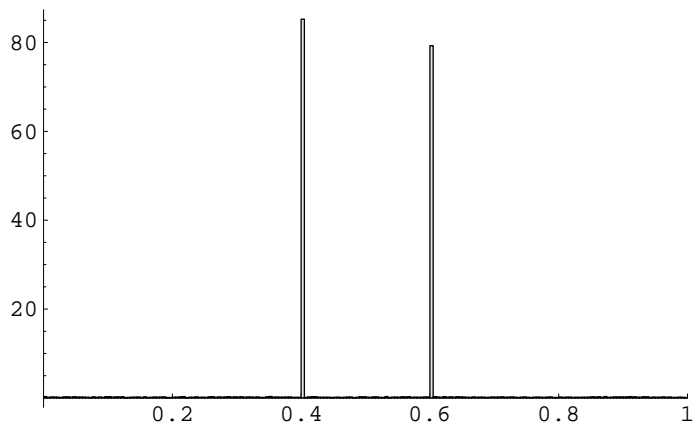}
\caption{\label{fig:finalLPhistogram}\textit{Histogram of the average distribution of opinions at $t=500$ when opinions evolve according to Rule 1. }}
\end{figure}

After 500 time steps, as shown in Figure~\ref{fig:finalLPhistogram}, most individuals have adopted either the opinion $\Omega=0.4$ or the opinion $1-\Omega=0.6$. The two peaks at 0.4 and 0.6 correspond, respectively, to $42.61 \%$ and $39.51 \%$ of the total population. These values include the percentages of opinion leaders (refer to Table~\ref{initialNumbers}). Note that opinion leaders with opinion 0.4, who are slightly more numerous, eventually convinced more people to adopt their opinion than the opinion $0.6$ of the other group of opinion leaders.
 
Except very few of them ($0.025 \%$), all  individuals belonging to group 1 (less aware) changed opinion. Those  who did not change, were found to have only one neighbor having a lower level of awareness than their own. According to Rule~1,  these individuals could not change opinion.

Highly committed individuals have a strong influence on the formation of opinion in societies evolving according to Rule 1. This rule leads to polarized societies divided into two consensual subgroups.

\subsection{Rule 2: Adopt weighted average with a probabilistic condition}

Starting from the same 20 initial populations of 1000 individuals we studied their temporal evolution assuming that an individual $i$ with level of awareness $s_i$ and opinion $\omega_1$ adopts the average opinion $\overline{\omega_i}$ of her neighborhood with a probability $(1-s_i)^4$.

\begin{figure}[h]
\centering\includegraphics[scale=0.9] {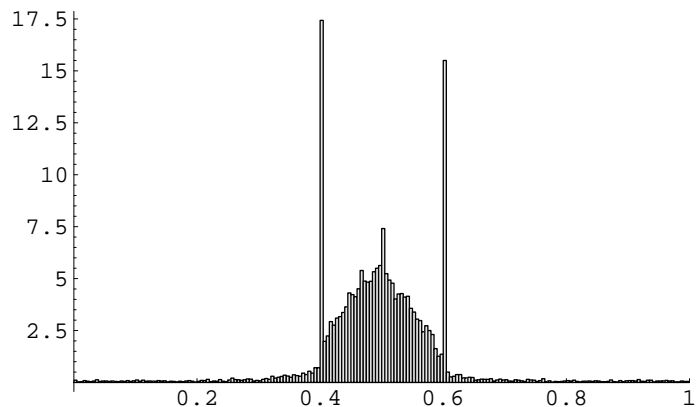}
\caption{\label{fig:finalWAPhistogram}\textit{Histogram of the distribution of opinions at $t=500$ when opinions evolve according to Rule~2.}}
\end{figure}

It is clear that if all individuals changing opinion adopt the weighted average opinion of their neighbors, they tend towards adopting opinions distributed around 0.5 as illustrated by the histogram (Figure~\ref{fig:finalWAPhistogram}). After 500 time steps, compared to what has been observed in the case of Rule~1,  even though more individuals did change opinion much less have finally adopted the opinions 0.4 and 0.6 promoted by the most influential individuals. As a matter of fact, only $3.14 \%$ and $2.7 \%$ changed opinion to respectively adopt opinions 0.4 and 0.6. 

As suggested above, we observed the formation of an important group ($74.07 \%$) of individuals having an opinion in the open interval $]0.4, 0.6[$ with a central peak at $0.5$ (see Figure~\ref{fig:finalWAPhistogram}). The percentages of individuals having an opinion either less than 0.4 or greater than 0.6 are only equal to $5.71 \%$ and $4.41 \%$ respectively. 

Compared to Rule~1 in which an individual adopts the opinion of her neighbor with the highest level of awareness if this level of awareness is higher than her own, in the case of Rule~2, an individual adopts the weighted average opinion of her neighbors with no restrictive condition. It is, therefore, not surprising to find that more individuals did change opinion during the evolution according to Rule~2 than during the evolution according to Rule~1. Actually all individuals of groups 1 and 2, that is, all individuals having a level of awareness less than 0.5, changed opinion. Moreover the vast majority of individuals belonging to group 3, that is, $99.05 \%$ of the total number of individuals belonging to this group, changed opinion.  
As expected, among the more aware individuals belonging to group 4, who are more difficult to convince, only $40.9 \%$ of them changed opinion. 

In order to verify the formation of a growing group of individuals having an opinion in the open interval $]0.4, 0.6[$ we analyzed the distribution of opinions after 1000 time steps and found that the percentage of individuals having an opinion in the open interval $]0.4, 0.6[$ slightly increased from $74.07 \%$ to $76.4 \%$ of the total population. The histogram of the distribution of opinions after 1000 time steps is very similar to the histogram obtained after 500 time steps. The temporal evolution is, therefore, rather slow and after 1000 time steps we are close to the steady state. Table~\ref{percentagesR2Unconvinced} gives, for each group, the percentages of individuals who did not change opinion after, respectively, 500 and 1000 time steps.  Note that these percentages refer to the number of individuals in each group.

\begin{table}[h]
 \begin{center}
 \begin{tabular}{|c|c|c|c|c|}
\hline
 time             & group 1 & group  2 & group  3 &  group 4\\ 
 \hline
 500 & 0 & 0  & 0.95 & 59.1 \\  
 1000 & 0 & 0 & 0.03 & 43.5 \\
\hline
\end{tabular}
\end{center}
\caption{\textit{Rule~2: percentages of the number of individuals in each group who did not change opinion after, respectively, 500 and 1000 time steps.}}
\label{percentagesR2Unconvinced}
\end{table}

\subsection{Rule 3: Adopt different weighted averages depending upon group membership with a probabilistic condition}

Considering the assumption that individuals with lower levels of awareness are less aware than individuals with higher levels of awareness, we suppose that individuals belonging to groups 1 and 2, that is, individuals with a level of awareness $s < 0.5$ may change their opinion to adopt the weighted average opinion of all their neighbors with a probability $(1-s)^4$, whereas individuals belonging to groups 3 and 4, that is, individuals with a level of awareness $s \geq 0.5$ may change their opinion to adopt the weighted average opinion of their neighbors having a higher level of awareness than their own with a probability $(1-s)^4$. An individual has a higher level of awareness than all her neighbors never changes opinion. This mixed rule tries to take into account the fact that less aware individuals do not have fixed positions on political issues whereas more aware individuals, having a more consistent ideology, may only be influenced by the elite. 

\begin{figure}[h]
\centering\includegraphics[scale=0.9] {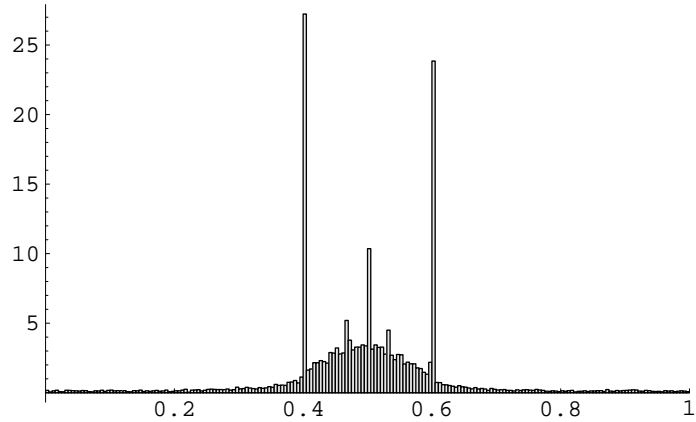}
\caption{\label{fig:finalMixed2neighborhood500histogram}\textit{Histogram of the distribution of opinions at time $t=500$ when the populations evolve according to mixed Rule~3.}}
\end{figure}

\begin{figure}[h]
\centering\includegraphics[scale=0.9] {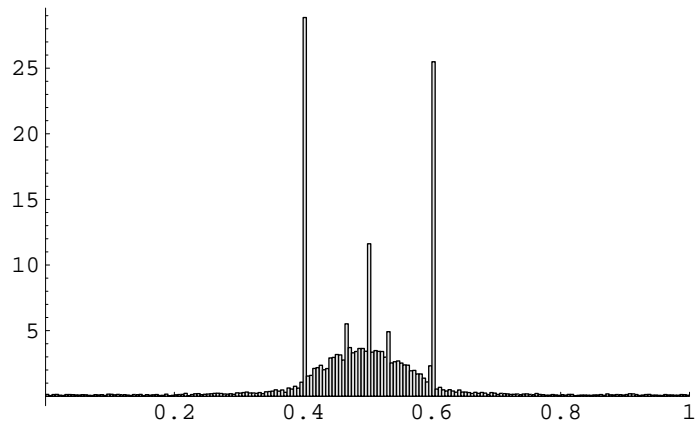}
\caption{\label{fig:finalMixed2neighborhood2000histogram}\textit{Histogram of the distribution of opinions at time $t=2000$ when the populations evolve according to mixed Rule~3.}}
\end{figure}

In Figure~\ref{fig:finalMixed2neighborhood500histogram}, representing the histogram of the average distribution of opinions after 500 time steps, we observe, as in the case of Rule~2, a relatively important but, however, less pronounced group (56.33 \% of the total population) of individuals having an opinion in the open interval $]0.4, 0.6[$ with a marked central peak at $0.5$ ($3.68 \%$ of the total population). Increasing the number of iterations up to 2000 confirms this feature 
(see Figure~\ref{fig:finalMixed2neighborhood2000histogram}). The temporal evolution is, here again, rather slow,  the central peak, for instance, grows from $3.68 \%$ to $4.15 \%$ of the total population when the number of time steps increases from 500 to 2000. The evolutions according to Rules 2 and 3 are quite similar.

Table~\ref{percentagesR3Unconvinced} gives, for each group, the percentages of individuals who did not change opinion after, respectively, 500 and 2000 time steps.  These percentages refer to the number of individuals in each group.

\begin{table}[h]
 \begin{center}
 \begin{tabular}{|c|c|c|c|c|}
\hline
 time             & group 1 & group  2 & group  3 &  group 4\\ 
 \hline
 500 & 0 & 0  & 13.72 & 68.09 \\  
 2000 & 0 & 0 & 13.02 & 46.32 \\
\hline
\end{tabular}
\end{center}
\caption{\textit{Rule~3: percentages of the number of individuals in each group who did not change opinion after, respectively, 500 and 2000 time steps.}}
\label{percentagesR3Unconvinced}
\end{table}

\subsection{Rule 4: Adopt Leader's Opinion with a Threshold Condition}

Starting from the same 20 initial populations we study their temporal evolution assuming that an individual with level of awareness $s$ and opinion $\omega$ adopts the opinion $\omega_{\rm leader}$ of her neighbor having the highest level of awareness $s_{\rm leader}$ if $s_{\rm leader} > s$ and the relation 
$|\omega-\omega_{\rm leader}|<(1-s)^4$ is satisfied.  

\begin{figure}[h]
\centering\includegraphics[scale=0.9] {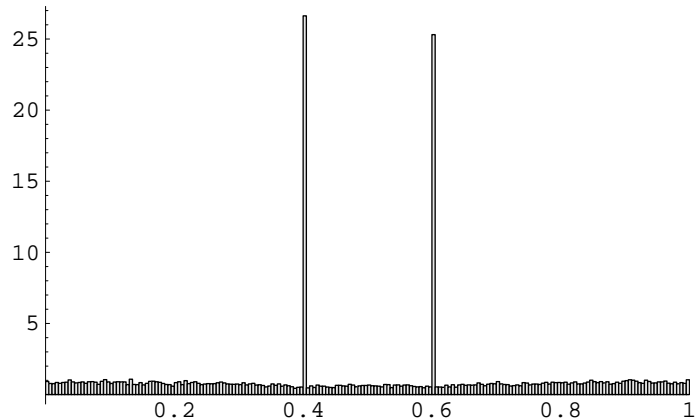}
\caption{\label{fig:finalLThistogram}\textit{Histogram of the distribution of opinions at time $t=500$ when the populations evolve according to Rule~3.}}
\end{figure}

Imposing a threshold condition and requiring that the individual to be convinced should have a level of awareness inferior to the level of awareness of the neighbor with the highest one slows down the evolution of the distribution of opinions as illustrated on Figure~\ref{fig:finalLThistogram}. Except for the two peaks at 0.4 and 0.6 that represent respectively, in this case, only $13.07 \%$ and $12.4 \%$ of the total population, the remaining part of the distribution is approximately uniform. The percentages of individuals in each group who have not changed opinion after 500 time steps are, respectively equal to  $17.99 \%$, $71.97 \%$,  $94.8 \%$, and $99.95 \%$. For the moderately aware and more aware individuals of groups 2, 3, and 4, these percentages are rather high.

\subsection{Rule 5: Adopt Weighted Average with a Threshold Condition}

Starting from the same 20 initial populations of 1000 individuals we study their temporal evolution assuming that an individual with level of awareness $s$ and opinion $\omega$ adopts the average opinion $\overline{\omega}$ of her neighborhood if the relation $|\omega-\overline{\omega}|<(1-s)^4$ is satisfied.

\begin{figure}[h]
\centering\includegraphics[scale=0.9] {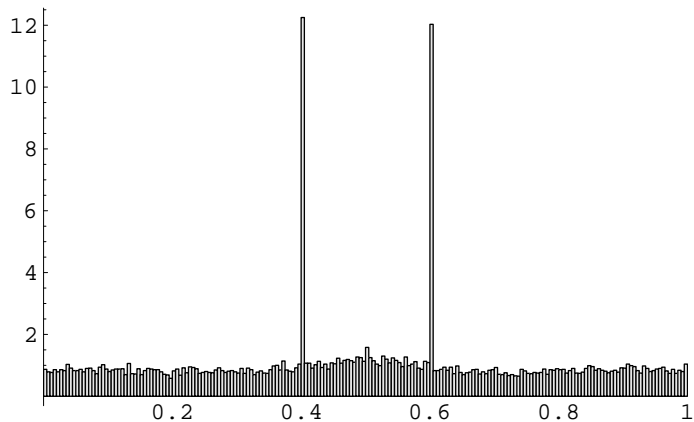}
\caption{\label{fig:finalWAThistogram}\textit{Histogram of the distribution of opinions at $t=500$ when opinions evolve according to Rule~5.}}
\end{figure}

Here again, as for Rule~4, we observe a slower evolution of the distribution of opinions according to Rule~5 when we impose a threshold condition instead of a probabilistic one (see Figure~\ref{fig:finalWAThistogram}). Much less individuals changed opinion compared to what was observed when the system evolved, following Rule~2 (see Table~\ref{percentagesR5Unconvinced}).
The two peaks at 0.4 and 0.6 represent just $5.68 \%$ and $5.58 \%$ respectively, that is, a mere $1.39 \%$ of the total population (excluding opinion leaders) has adopted one of the two opinions promoted by opinion leaders. The central group of the individuals with an opinion in the open interval $]0.4, 0.6[$ is also much less evident, only $22.02 \%$ of the total population compared to the $74.07 \%$ when the system evolved following Rule~2.

\begin{table}[h]
 \begin{center}
 \begin{tabular}{|c|c|c|c|c|}
\hline
 time             & group 1 & group  2 & group  3 &  group 4\\ 
 \hline
 500 & 16.92 & 69.86  & 93.81 &  99.73\\  
\hline
\end{tabular}
\end{center}
\caption{\textit{Rule~5: percentages of the number of individuals in each group who did not change opinion after 500 time steps.}}
\label{percentagesR5Unconvinced}
\end{table}

As we did in the case of Rule~2, we increased the number of time steps to see if the hardly visible small bump centered around opinion 0.5 in the histogram of the distribution of opinions (Figure~\ref{fig:finalWAThistogram}) would grow. Actually nothing changed, we reobtain exactly the same numbers, that is, after 500 time steps the system had already reached a steady state. 

\subsection{Rule 6: Adopt different weighted averages depending upon group membership with a threshold condition}

Starting from the same 20 initial populations of 1000 individuals we study their temporal evolution assuming that an individual with level of awareness $s < 0.5$ and an opinion $\omega$ may change her opinion to adopt the weighted average opinion $\overline{\omega}$ of all her neighbors if the condition $|\omega-\overline{\omega}| < (1-s)^4$ is satisfied, whereas an individual with level of awareness $s \geq 0.5$ and an opinion $\omega$ may change her opinion to adopt the weighted average opinion 
$\overline{\omega}_{\rm leaders}$ of their neighbors having a higher level of awareness than her own if the condition $|\omega-\overline{\omega}_{\rm leaders}| < (1-s)^4$ is satisfied.

\begin{figure}[h]
\centering\includegraphics[scale=0.9] {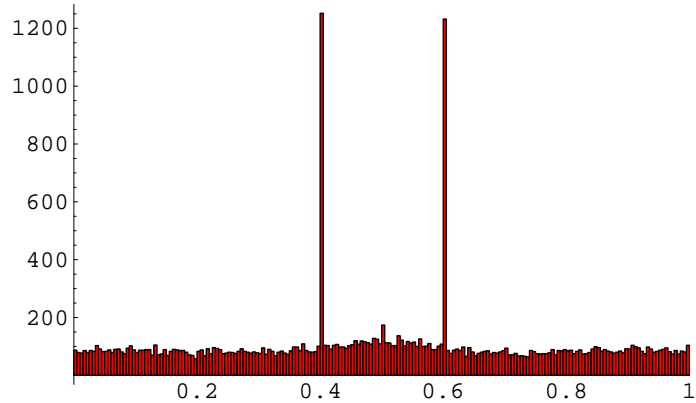}
\caption{\label{fig:finalMixed2neighborhood500withThresholdhistogram}\textit{Histogram of the distribution of opinions at $t=500$ when opinions evolve according to Rule~6.}}
\end{figure}

As already observed for Rules 4 and 5, imposing a threshold condition slows down the evolution and generate a small central group of individuals with opinions around 0.5 as illustrated in 
Figure~\ref{fig:finalMixed2neighborhood500withThresholdhistogram} and 
Table~\ref{percentagesR6Unconvinced}.

\begin{table}[h]
 \begin{center}
 \begin{tabular}{|c|c|c|c|c|}
\hline
 time             & group 1 & group  2 & group  3 &  group 4\\ 
 \hline
 500 & 17 & 69.86  & 94.91 &  100\\  
\hline
\end{tabular}
\end{center}
\caption{\textit{Rule~6: percentages of the number of individuals in each group who did not change opinion after 500 time steps.}}
\label{percentagesR6Unconvinced}
\end{table}

The evolution of the distribution of opinions according to Rules 5 and 6 are very similar (compare results given in Tables~\ref{percentagesR5Unconvinced} and \ref{percentagesR6Unconvinced}). Actually, the individuals changing opinions belong essentially to groups 1 and 2, and these individuals evolve exactly in the same manner for both Rules 5 and 6.

\section{Conclusion}
We have studied the evolution of the distribution of opinions of 20 different 1000-individual societies in which each individual, located at a vertex of a directed graph, is characterized by her name (an integer in the range $[1, 1000]$), the list of people she is interacting with, her level of awareness (a real $s$ between 0 and 1), and her opinion (a real $\omega$ between 0 and 1). The out-degree of a vertex, representing the number of persons who may influence the opinion of the individual located at that vertex, is a random number uniformly distributed between 1 and $n_{\max}=7$. The in-degree of a vertex, representing the number of persons who may be influenced by the opinion of the individual located at that vertex, is a random number Pareto distributed with a minimum value parameter $x_0=1$ and a shape parameter $\sigma=2$. The resulting power-law behavior of the in-degrees implies that there exist individuals who may influence a large number of other individuals. The group $rN = 100$ individuals having the highest in-degrees, is assigned a level of awareness $s = 1$ and, with a probability $\frac{1}{2}$, an opinion either equal to $\Omega=0.4$ or $1-\Omega=0.6$.  This subnetwork of the most influential individuals are opinion leaders who never change opinion during temporal evolutions of the system.  
The evolution rules are of three different types. An individual can either adopt the opinion of the individual, among those she is connected to, who has the highest level of awareness if this level of awareness is higher than her own or adopt the weighted average of the opinions of all the individuals she is connected to or adopt the weighted average of the opinions of the individuals having a higher level of awareness than her own among those she is connected to. Each of these rules has two versions. An individual with a level of awareness $s$ may either adopt a new opinion with a probability equal to  $(1-s)^4$ or adopt a new opinion if this opinion differs from her own by less than $(1-s)^4$. The choice of the power $4$ makes individuals with a level of awareness less than $0.25$ very easy to convince and individuals with a level of awareness greater than $0.75$ extremely reluctant to adopt a new opinion.

All numerical simulations have shown that the temporal evolution according to Rule 1 (which consists, for a person with level of awareness $s$, to adopt, with a probability $(1-s)^4$, the opinion of the individual having the highest level of awareness among those she is connected to when the leader's level of awareness is higher than her own) leads very quickly to the formation of two important, roughly equivalent, polarized consensual subgroups of individuals having either the opinion 0.4 or the opinion 0.6, whereas the temporal evolution according to Rule 5 (which consists, for a person with level of awareness $s$ and an opinion $\omega$, to adopt the weighted average opinion 
$\overline{\omega}$ of the individuals she is connected to if the condition $|\omega-\overline{\omega}|<(1-s)^4$ is verified) leads rather slowly to a much less polarized society with the formation of a group of individuals having a bell-shaped distribution of opinions centered at $0.5$ accompanied by less pronounced peaks corresponding to the two groups of opinion leaders. 

The formation of polarized consensual subgroups seems therefore to reveal that people are essentially influenced by opinion leaders (elite discourse) conveyed through the media by, for example, politicians and journalists, whereas the formation of a group with opinions distributed around a median opinion appears to be the result of interactions with closely related individuals such as family members, friends and coworkers.

\end{document}